\documentclass[12pt]{article}
\usepackage{amssymb}
\usepackage{amsfonts}
\usepackage{amsmath}
\usepackage{graphicx}
\usepackage{color}

\def\Z{\mathbb Z}

\def\R{\mathbb R}
\def\NN{\mathbb N}

\newcommand{\De}{\Delta}

\newcommand{\tH}{\widetilde{H}}
\newcommand{\tI}{\widetilde{I}}

\newcommand{\hI}{\widehat{I}}
\newcommand{\hM}{\widehat{M}}
\newcommand{\cD}{{\cal D}}
\newcommand{\hcD}{\widehat{\cal D}}
\newcommand{\cG}{{\cal G}}
\newcommand{\cR}{{\cal R}}

\renewcommand{\=}{\ =\ }
\newcommand{\und}{\qquad\text{and}\qquad}
\newcommand{\+}{{\dagger}}
\newcommand{\sfrac}[2]{{\textstyle\frac{#1}{#2}}}
\newcommand{\pa}{\partial}

\newcommand{\res}{{\mathrm{res}}}
\newcommand{\tr}{{\mathrm{tr}}}
\newcommand{\im}{{\mathrm{i}}}

\newcommand{\ph}{\phantom{-}}

\newcommand{\beq}{\begin{equation}}
\newcommand{\eeq}{\end{equation}}
\newcommand{\bea}{\begin{eqnarray}}
\newcommand{\eea}{\end{eqnarray}}

\advance \textheight by \topskip
\makeatletter
\@addtoreset{equation}{section}

\makeatother
\def\beq{\begin{equation}}
\def\eq{\end{equation}}

\def\12{\frac{1}{2}}

\begin{document}

\begin{titlepage}
\setcounter{page}{0}
\begin{flushright}
ITP--UH--21/13\\
\end{flushright}

\vskip 1.5cm

\begin{center}

{\LARGE\bf
Nonlinear supersymmetry \\[8pt] in the quantum Calogero model
}

\vspace{12mm}
{\Large Francisco Correa${}^\+$, 
Olaf Lechtenfeld${}^{\times\circ}$ and 
Mikhail Plyushchay${}^*$}\\[8mm]

\noindent ${}^\+${\em 
Centro de Estudios Cient\'ificos (CECs)\\
Casilla 1468, Valdivia, Chile } \\
{Email: correa@cecs.cl}\\[6mm]

\noindent ${}^\times${\em
Institut f\"ur Theoretische Physik and Riemann Center for Geometry and Physics\\
Leibniz Universit\"at Hannover \\
Appelstra\ss{}e 2, 30167 Hannover, Germany }\\
{Email: lechtenf@itp.uni-hannover.de}\\[6mm]

\noindent ${}^{\circ}${\em
Centre for Quantum Engineering and Space-Time Research\\
Leibniz Universit\"at Hannover \\
Welfengarten 1, 30167 Hannover, Germany }\\[6mm]

\noindent ${}^*${\em 
Departamento de F\'isica, Universidad de Santiago de Chile\\
Casilla 307, Santiago 2, Chile } \\
{Email: mikhail.plyushchay@usach.cl}

\vspace{12mm}

\begin{abstract}
\noindent
It is long known that the rational Calogero model describing $n$~identical particles
on a line with inverse-square mutual interaction potential is quantum superintegrable.
We review the (nonlinear) algebra of the conserved quantum charges and the intertwiners
which relate the Liouville charges at couplings $g$ and~$g{\pm}1$. For integer
values of~$g$, these intertwiners give rise to additional conserved charges commuting
with all Liouville charges and known since the 1990s. We give a direct construction of
such a charge, the unique one being totally antisymmetric under particle permutations.
It is of order $\sfrac12n(n{-}1)(2g{-}1)$ in the momenta and squares to a polynomial in 
the Liouville charges. With a natural $\Z_2$ grading, this charge extends the algebra of 
conserved charges to a nonlinear supersymmetric one. We provide explicit expressions for 
intertwiners, charges and their algebra in the cases of two, three and four particles.
\end{abstract}

\end{center}

\end{titlepage}

\section{Introduction and summary}

The Calogero model \cite{Cal71} and its generalizations are the workhorses of
multi-particle integrable systems
(for reviews, see~\cite{OlshaPere81-rev,OlshaPere83-rev,Poly1,Poly06-rev}).
The $n$-particle system is well known to be not only classically and quantum integrable 
but superintegrable~\cite{Woj}, meaning that, on top of the $n$~mutually commuting 
Liouville integrals, one has $n{-}1$ additional algebraically independent constants 
of motion which form a quadratic algebra~\cite{Kuz,BarReg77}.\footnote{
In this paper we only deal with the $A_{n-1}$ model, including for convenience
the additional $A_1$ part representing the center of mass.
Generalizations to all Coxeter root systems are straightforward.}
Furthermore, when the real-valued coupling constant~$g$~
is an integer,\footnote{
In the mathematical literature, $g{-}1=m$ is called the `multiplicity'.} 
the model possesses even more structure, which has been termed 
`analytically integrable'~\cite{Krichever}.
In this case, there exist additional conserved charges commuting with all Liouville charges,
which are not algebraic combinations of them. The corresponding ring of commuting
differential operators has been termed `supercomplete'.

For one degree of freedom (corresponding to $n{=}2$ after separating the center of mass), 
the occurrence of odd-order (ordinary) differential operators 
commuting with the Hamiltonian is well understood for a long 
time~\cite{BurCha,BC2,Baker,Ince} and is tied to the algebro-geometric, or `finite-gap', 
nature of the Hamiltonian~\cite{Krich78,Brezh}.
Progress for more degrees of freedom started in 1990 with the first examples of two-dimensional
($n{=}3$) finite-gap Schr\"odinger operators and the construction of the intertwiners
(shift operators) at multiplicity~$m{=}1$ ($g{=}2$)~\cite{ChaVes90}.
Subsequent works extended these results to larger particle numbers ($n{>}3$) and higher multiplicity
($g{>}2$), focussing mostly on the Baker--Akhiezer function~\cite{ChaStyVes93}--\cite{Chalykh08-rev}.
An explicit formula for the additional conserved charges was given by Berest~\cite{Berest98} 
(reproduced in~\cite{ChaFeiVes99}). Furthermore, Chalykh~\cite{Chalykh96} showed how these
charges are obtained by Darboux dressing with intertwiners, 
with concrete examples for $n{=}3$ and $g{=}2$. 
An important parallel development was the application of Dunkl operators~\cite{DunklXu} 
to construct intertwining operators for higher multiplicity~\cite{Opdam,Heckman}.

In this paper we combine the technology of Dunkl operators with the Darboux dressing method
to provide another explicit construction of the additional conserved charges for arbitrary
multiplicity (any integer~$g$). As physicists, we are mainly interested in observables 
totally symmetric or antisymmetric under particle exchange.
For a discussion of this issue, see \cite{Poly1,Poly06-rev}.
There exists a unique totally antisymmetric conserved charge~$Q$ which factorizes into $2g{-}1$
intertwiners like in the one-dimensional case. Each intertwiner is a totally antisymmetric
differential operator of order~$\sfrac12n(n{-}1)$. 
The new charge $Q$ squares to the $(2g{-}1)$th power of a universal polynomial in the Liouville charges. 
Adding just $Q$ to the quadratic algebra of the
Liouville and non-Liouville integrals seems to produce no further independent integrals.
Employing any two-particle exchange as a $\Z_2$ grading, the enhanced algebra acquires
a (nonlinear) supersymmetry structure.  In fact, the formulae (\ref{algebra}) and (\ref{allpoly}) 
represent the key results of the present paper.

After reviewing the conserved integrals and their algebra for generic values of the coupling,
we provide the construction of~$Q$ with the help of the Heckman--Opdam intertwiners and
Darboux dressing, compute its square and present the nonlinear supersymmetry structure.
Finally, explicit formulae are given for the cases of two, three and four particles.
Separating the center of mass, one may always reduce the number of degrees of freedom by one.
In addition, the conformal invariance allows one to further separate the reduced system into
a radial and an angular part (in spherical coordinates). This decomposes the three-particle system
into its center of mass, a particle on the half-line in an $r^{-2}$ potential and a particle on a circle
in a P\"oschl-Teller potential (see, e.g.~\cite{FeLePo13}).
Reducing the four-particle rational Calogero system in this manner, the angular subsystem describes
a particle on the two-sphere in a cuboctahedric Higgs oscillator potential~\cite{HaNeYe08}.
Hence, the four-particle case is the simplest one not separable into one-dimensional systems and
is firstly worked out in this paper.

\section{Integrals of the Calogero model for generic coupling}

The quantum phase space of the $n$-particle Calogero model, 
defined by the Hamiltonian 
\begin{equation} \label{calham}
H\=\sfrac12\sum_i p_i^2\ +\ \sum_{i<j}\frac{g(g{-}1)}{(x^i{-}x^j)^2}\ ,
\end{equation} 
is spanned by the particle coordinates~$x^i$ and their conjugate 
momenta~$p_j$, with $i,j=1,2,\ldots,n$.
For convenience, we take $\hbar=m=1$, so $[x^i, p_j]=\im\delta^i_j$,
and the only parameter is the Calogero coupling~$g\in\R$.
We shall often represent momenta by partial derivatives,
\begin{equation}
p_j \= -\im \sfrac{\pa}{\pa x^j} \= -\im\pa_j\ .
\end{equation}
It is also useful to introduce the center-of-mass momentum and coordinate,
\begin{equation} \label{calmom}
P\=\sum_{i=1}^n p_i \und X\=\frac1n\sum_{i=1}^n x^i\ .
\end{equation}

The standard Liouville contants of motion $I_k$ for any value of~$g$
can be constructed from powers of the Dunkl operators~\cite{DunklXu}
\begin{equation}
\pi_i \= p_i + \im\sum_{j(\neq i)} \frac{g}{x^i{-}x^j}s_{ij} 
\qquad\Leftrightarrow\qquad
\cD_i \= \pa_i - \sum_{j(\neq i)} \frac{g}{x^i{-}x^j}s_{ij}\ ,
\end{equation}
where $s_{ij}=s_{ji}$ are the two-particle permutation operators, satisfying 
$s_{ij}x^i=x^js_{ij}$,  $s_{ij}\pa_i=\pa_js_{ij}$ and  $s_{ij}^2=1$.

The independent Liouville integrals read
\begin{equation}
I_k \= \res\Bigl(\sum_i \pi_i^k\Bigr) \qquad\textrm{for}\quad k=1,2,\ldots,n\ ,
\end{equation}
with $\res(A)$ denoting the restriction of an operator~$A$ to the subspace of 
states which are totally symmetric under any two-particle exchange. 
Because the Dunkl operators commute, $[\pi_i,\pi_j]=0$,
it is easy to prove that the $I_k$ commute with one another,
\begin{equation} \label{Icom}
[I_k,I_\ell]\=0\ .
\end{equation} 
The first three Liouville integrals read
\begin{equation}
I_1\=P\ ,\qquad
I_2\=2H\ ,\qquad
I_3\=\sum_i p_i^3\ +\ 3\sum_{i<j}\frac{g(g{-}1)}{(x^i{-}x^j)^2}(p_i{+}p_j)\ ,
\end{equation}
where $P$ and $H$ are given in (\ref{calmom}) and (\ref{calham}), respectively.
It can be useful to transfer the Hamiltonian (\ref{calham}) to a `potential-free' 
frame by means of a similarity transformation, see Appendix~A.
Since (\ref{Icom}) contains in particular $[H,I_k]=0$,  
the $I_k$ form $n$ involutive constants of motion,
whose leading term for large $|x^i{-}x^j|$ is $\sum_i p_i^k$.
Another important fact, to be used later, is the equality
\begin{equation}
I_k(g) \= I_k(1{-}g)\ ,
\end{equation}
which follows because the $I_k$ depend only on 
the combination~$g(g{-}1)$.

Together with
\begin{equation}
D\=\sfrac12\sum_i(x^ip_i+p_ix^i) \und
K\=\sfrac12\sum_i (x^i)^2\ ,
\end{equation}
the Hamiltonian is part of an $sl(2)$ algebra,
\begin{equation}
[D,H]\=2\im H\ ,\qquad
[D,K]\=-2\im K\ ,\qquad
[K,H]\=\im D\ .
\end{equation}
This fact allows for the construction of many additional integrals,
from which we may choose $n{-}1$ functionally independent ones.
In the following, we derive their structure and algebra,
roughly following \cite{Woj} and~\cite{Kuz} (see also~\cite{BarReg77}).

Firstly, observe that the $I_k$ have a definite scaling dimension,
\begin{equation}
\sfrac1\im[D,I_k]\=k I_k\ .
\end{equation}
Secondly, we employ the conformal generator $K$ 
to create a new operator from each $I_k$,
\begin{equation}
\sfrac1\im[K,I_\ell]\ =:\ \ell J_\ell\ ,
\end{equation}
in particular
\begin{equation}
J_1\=nX\ ,\qquad J_2\=D\ ,\qquad
J_3\=\sfrac12\sum_i\bigl(x^i p_i^2+p_i^2 x^i\bigr)
+g(g{-}1)\sum_{i<j}\frac{x^i{+}x^j}{(x^i{-}x^j)^2}\ .
\end{equation}
It is straightforward to derive
\begin{equation}
\sfrac1\im[D,J_\ell]\=(\ell{-}2)J_\ell \und
\sfrac1\im[H,J_\ell]\=-I_\ell\ .
\end{equation}
The Jacobi identity implies that $\bigl[H,[I_\ell,J_k]\bigr]=0$, 
so $[I_\ell,J_k]$ must be a linear combination of the $I_m$,
\begin{equation}
\sfrac{\im}{k}[I_k,J_\ell] \= I_{k+\ell-2} \= \sfrac{\im}{\ell}[I_\ell,J_k]\ ,
\end{equation}
in particular
\begin{equation}
\im[I_1,J_\ell]\=I_{\ell-1} \und
\im[I_2,J_\ell]\=2\,I_\ell\ .
\end{equation}
We note that the $J_m$ form a closed algebra,
\begin{equation}
\im[J_k,J_\ell] \=(k{-}\ell)J_{k+\ell-2}\ ,
\end{equation}
thus the shifted operators $L_k=J_{k+2}$ satisfy the Witt algebra.
They are not invariant, but their simple time evolution, 
$\dot{J}_\ell=\im[H,J_\ell]=I_\ell$,
suggests a composite of $I_k$ and $J_\ell$ for further integrals.

Thus, thirdly, we define the symmetrized antisymmetric combinations
\begin{equation}
L_{k,\ell}\=\sfrac12 (I_k J_\ell+J_\ell I_k) - \sfrac12 (I_\ell J_k+J_k I_\ell)
\ \equiv\ \sfrac12 \{I_k, J_\ell\} -  \sfrac12 \{I_\ell, J_k\} \= -L_{\ell,k}\ ,
\end{equation}
which indeed all commute with the Hamiltonian, $[H,L_{k,\ell}]=0$.
Note that we have moved from the Lie algebra to the universal enveloping algebra.
The $L_{k,\ell}$ have scaling dimension~$k{+}\ell{-}2$ and begin with a term
of the form $x\,p^{k+\ell-1}$ in leading order.
However, they form an overcomplete set of integrals.
Two interesting choices of functionally independent ones are
\begin{equation}\label{eql}
L_{1,\ell}\ =:\ E_\ell\= \sfrac12\{P, J_\ell\} - \sfrac12 \{I_\ell, nX\} \und 
L_{2,\ell}\ =:\ F_\ell\= \{H, J_\ell\} - \sfrac12 \{I_\ell, D\}
\end{equation}
with $\ell=1,2,\ldots,n$, but note that 
\begin{equation}
E_1\ \equiv\ 0\ \equiv F_2 \qquad\textrm{while}\qquad
F_1 \= -E_2 \= \{H, nX\} - \sfrac12\{P, D\} \ .
\end{equation}

For our purposes, the $F_\ell$ are more suitable, because they bear
a close relation to the Casimir element of the $sl(2)$ algebra,
\begin{equation}
C \= KH+HK-\sfrac12 D^2\ .
\end{equation}
Namely, the latter generates the $F_\ell$ directly from the $I_\ell$,
\begin{equation}
\sfrac1{\im\ell}[C,I_\ell]\= F_\ell\ .
\end{equation}
A slightly more convenient form of these integrals is
\begin{equation}
F_\ell \= 2J_\ell H-\bigl(D-\sfrac{\im}{2}(\ell{-}2)\bigr)I_\ell
\= 2H J_\ell-I_\ell\bigl(D+\sfrac{\im}{2}(\ell{-}2)\bigr)\ .
\end{equation}
Hence, a useful set of $2n{-}1$ constants of motion for the $n$-particle Calogero model 
at arbitrary coupling~$g$ is $\{P,H,I_3,\ldots,I_n,F_1,F_3,\ldots,F_n\}$.
Their quadratic algebra reads
\begin{eqnarray} \label{quad1}
&\im[I_k,I_\ell]\=0\ ,\qquad
\sfrac{\im}{k}[I_k,F_\ell]\=I_{k+\ell-2}I_2-I_k I_\ell \= \sfrac{\im}{\ell}[I_\ell,F_k]\ ,\\[6pt]
&\begin{aligned} \label{quad2}
\im[F_k,F_\ell]&\= (k{-}\ell)\sfrac12\{F_{k+\ell-2},I_2\}
-(k{-}2)\sfrac12\{F_k,I_\ell\}+(\ell{-}2)\sfrac12\{F_\ell,I_k\} \\[4pt]
&\=(k{-}\ell)(F_{k+\ell-2}{+}\im I_{k+\ell-2})I_2
-(k{-}2)(F_k{+}\im I_k)I_\ell+(\ell{-}2)(F_\ell{+}\im I_\ell)I_k \ .
\end{aligned}
\end{eqnarray}
In particular, one has
\begin{equation}
\im[I_1,F_\ell]\=I_{\ell-1}I_2-I_\ell I_1 \= \sfrac{\im}{\ell}[I_\ell,F_1]
\end{equation}
while $[I_2,F_\ell]=[I_\ell,F_2]=0$ trivially and $I_0\equiv n\mathrm{1}$.

Since only the first $n$ of the Liouville charges $I_k$ are independent, 
one should express $I_{n+1},\ldots,I_{2n-2}$ in terms of them. 
This can be done via formulae collected in Appendix~\ref{apb}. 
Inserting the corresponding expressions
into (\ref{quad1}) and~(\ref{quad2}) changes the quadratic algebra to a polynomial one of order $2n{-}1$.

\section{Integrals of the Calogero model for integer coupling}

It is known that, for integer values of the coupling~$g$, 
the $A_{n-1}$ Calogero model becomes `algebraically integrable'
\cite{ChaVes90}--\cite{Chalykh08-rev},
meaning that the number of functionally independent integrals of motion increases
from $2n{-}1$ to $2n$. In other words, there exists one further algebraically 
independent constant of motion, which we call~$Q$. Its construction relies on the
existence of intertwining operators which relate the integrals $I_k$ at couplings
$g$ and~$g{+}1$. It allows not only for the explicit evaluation of the energy 
eigenstates from the free-particle ones, but, together with the 
$g\leftrightarrow1{-}g$ symmetry, also for the construction of~$Q$, in involution
with all~$I_k$.

It was shown by Heckman~\cite{Heckman} (see also~\cite{Opdam}) that
\begin{eqnarray}
M(g)\,I_k(g) \= I_k(g{+}1)\,M(g) &&\quad\textrm{for}\qquad
M(g)\=\res\Bigl(\prod_{i<j}(\cD_i{-}\cD_j)(g)\Bigr)\ ,
\label{up}\\ \label{down}
M(g)^*\,I_k(g{+}1) \= I_k(g)\,M(g)^* &&\quad\textrm{for}\qquad
M(g)^*\=\res\Bigl(\prod_{i<j}(\cD_i{-}\cD_j)(-g)\Bigr)\ ,
\end{eqnarray}
consistent with $M(g)^*=M(-g)$, due to the $g\leftrightarrow1{-}g$ symmetry.
Hence, we may rewrite (\ref{down}) in another suggestive form,
\begin{equation}
M(1{-}g)\,I_k(g) \= I_k(g{-}1)\,M(1{-}g)\ .
\end{equation}
The intertwiner $M(g)$ is totally antisymmetric in the $x^i$ variables
and is a differential operator of order $\sfrac12n(n{-}1)$.
An immediate consequence of (\ref{up}) and (\ref{down}) are the relations
\begin{equation}
\bigl[ M(g)^* M(g),I_k(g)\bigr]\=0 \und \bigl[ M(g)\,M(g)^*,I_k(g{+}1)\bigr]\=0\ .
\end{equation}
However, $M^*(g)\,M(g)$ or $M(g)\,M^*(g)$ do not constitute a new integral of motion, because
they are polynomials in the Liouville integrals,
\begin{equation} \label{Rdef}
M(g)^* M(g)\=M(-g)\,M(-g)^*\=\cR\bigl(I(g)\bigr)\ =:\ \cR(g)\ ,
\end{equation}
where we abbreviate $I=\{I_k\}$ and the like.
Since the coefficients of the polynomial~$\cR(I)$ do not depend on~$g$, 
they can be computed from the free case ($g{=}0$),
\begin{equation} \label{vdm}
\begin{aligned}
\cR(0) &\= M(0)^* M(0) \= \prod_{i<j}(\pa_i{-}\pa_j)^2 \\
& \= \left| \begin{matrix} 
1 & \pa_1 & \ldots & \pa_1^{n-1} \\
1 & \pa_2 & \ldots & \pa_2^{n-1} \\
\vdots & \vdots &  & \vdots \\
1 & \pa_n & \ldots & \pa_n^{n-1} 
\end{matrix} \right|^{\,\textstyle 2}
\= \left| \begin{matrix} 
1 & 1 & \ldots & 1 \\
\pa_1 & \pa_2 & \ldots & \pa_n \\
\vdots & \vdots &  & \vdots \\
\pa_1^{n-1}\!\! & \!\!\pa_2^{n-1}\!\!\! & \ldots & \!\!\!\pa_n^{n-1}
\end{matrix} \right| \cdot
\left| \begin{matrix} 
1 & \pa_1 & \ldots & \pa_1^{n-1} \\
1 & \pa_2 & \ldots & \pa_2^{n-1} \\
\vdots & \vdots &  & \vdots \\
1 & \pa_n & \ldots & \pa_n^{n-1} 
\end{matrix} \right| \\[4pt]
&\= \biggl| \Bigl( \sum_k \pa_k^{i+j-2} \Bigr)_{ij} \biggr|
\= (-1)^{\lfloor\sfrac{n}{2}\rfloor} \det\bigl(I_{i+j-2}(0)\bigr)_{ij}\ .
\end{aligned}
\end{equation}
It follows that
\begin{equation}
\cR(g) \= (-1)^{\lfloor\sfrac{n}{2}\rfloor} \det\bigl(I_{i+j-2}(g)\bigr)_{ij} 
\qquad\textrm{with}\quad I_0=n\ ,
\end{equation}
yielding a specific polynomial in $I_1, I_2, \ldots, I_{2n-2}$,
which is then reduced further by employing~(\ref{dependent}).
For a small number of particles, one obtains
\begin{eqnarray}
n=2: &&\!\! \cR = I_1^2 - 2 I_2 \ ,\\[4pt]
n=3: &&\!\! \cR = I_1^2 I_4 -2 I_1 I_2 I_3 + I_2^3 - 3 I_2 I_4 + 3 I_3^2 \nonumber\\ \label{poly3}
&&\!\!\quad = \sfrac16 \bigl( 
I_1^6-9I_1^4I_2+8I_1^3I_3+21I_1^2I_2^2-36I_1I_2I_3-3I_2^3+18I_3^2\bigr)\ ,\\[4pt] 
n=4: &&\!\! \cR = -I_{2226}+2I_{2235}+I_{2244}-3I_{2334}+4I_{246} 
             -4I_{255}+I_{3333}-4I_{336}+8I_{345}-4I_{444} \nonumber \\
&&\!\!\quad = \sfrac1{72} \bigl( 
9I_2^6-90I_2^4I_4-68I_2^3I_3^2+288I_2^2I_4^2+144I_2I_3^2I_4-24I_3^4-288I_4^3
\bigr)\ ,
\end{eqnarray}
where in the $n{=}4$ case we put $I_1=0$ and abbreviated 
$I_{k\ell m\dots}\equiv I_kI_\ell I_m\cdots$ for simplicity and shortness.

So far, we did not specialize the value of~$g$. 
It is clear that something special happens for $g\in\NN$.
In this case, a product of intertwiners relates the integrals $I_k(g)$ to the free ones, 
$I_k(0)=I_k(1)$,
\begin{equation} \label{Ldef}
L(g) I_k(1)\= I_k(g) L(g) \qquad\textrm{with}\qquad
L(g)\=M(g{-}1)M(g{-}2)\cdots M(2) M(1)
\end{equation}
for $g\ge2$, and the conjugate reads
\begin{equation}
L(g)^*\=M(-1)\cdots M(2{-}g)M(1{-}g) \, .
\end{equation}
From~(\ref{Rdef}) we learn that
\begin{equation}
L(g)\,L(g)^*\=\bigl({\cal R}(g)\bigr)^{g-1} \und L(g)^* L(g)\=\bigl({\cal R}(g{+}1)\bigr)^{g-1}
\end{equation}
commute with $I_k(g)$ and $I_k(g{+}1)$, respectively.

Suppose now that, in the free case, some operator $G(1)$ commutes 
with one of the Liouville charges,
\begin{equation}
\bigl[ G(1), I_k(1) \bigr] \= 0\ .
\end{equation}
A quick computation shows that its Darboux-dressed variant then commutes with the
corresponding charge at any integer coupling~\cite{CorYakPly08,PlyNie10,AraGuiPly12} 
\begin{equation}
\bigl[ G(g),I_k(g)\bigr]\=0 \qquad\textrm{for}\qquad
G(g)\=L(g) G(1) L(g)^*\ .
\end{equation}
This is consistent with the involution of the Liouville charges, due to
\begin{equation}
L(g) I_k(1) L(g)^* \= \bigl({\cal R}(g)\bigr)^{g-1} I_k(g)\ .
\end{equation}
There is a large choice for `naked' operators $G(1)$: Any polynomial in $\pa_i$
with constant coefficients will produce a conserved charge in involution with
all Liouville integrals~\cite{Chalykh96}. However, as physicists we are
interested in observables which are either totally symmetric or antisymmetric
under particle exchange, see \cite{Poly1,Poly06-rev}. 
The totally symmetric ones are already spanned by the
Liouville integrals. Hence, we are considering here only $G(1)$ which are
antisymmetric under any permutation~$s_{ij}$ of two particle labels.
The simplest such expression is
\begin{equation}
G(1) \= M(0) \= \prod_{i<j}(\pa_i{-}\pa_j)\ .
\end{equation}
Applying the Darboux dressing to this operator, we obtain~\footnote{
We may extend the definitions by declaring $L(1)=1$ and $Q(1)=M(0)$.}
\begin{equation}
Q(g)\= L(g)M(0)L(g)^* \=
M(g{-}1)M(g{-}2)\cdots M(1)M(0)M(-1)\cdots M(2{-}g)M(1{-}g)\ ,
\end{equation}
which builds a chain relating $I_k(g)=I_k(1{-}g)$ back to $I_k(g)$,
\begin{equation}
Q(g) I_k(1{-}g) \= I_k(g) Q(g) \qquad\Rightarrow\qquad
\bigl[ Q(g),I_k(g)\bigr] \=0\ .
\end{equation}
We have thus identified another conserved charge, which cannot be reduced to the set
$\{I_k,F_\ell\}$ and is of order $\sfrac12n(n{-}1)(2g{-}1)$.\footnote{
Another explicit formula is 
$Q(g)\sim\bigl(\textrm{ad}H(g)\bigr)^{\frac12n(n-1)(2g-1)} \Delta^{2g-1}$
with $\Delta=\prod_{i<j} (x^i-x^j)$ \cite{Berest98,ChaFeiVes99}.}

A second possibility to form such a chain starts from half-integer values of~$g$.
The candidate for conserved  charge then has the form
\begin{equation}
M(g{-}1)M(g{-}2)\cdots M(\sfrac32)M(\sfrac12)M(-\sfrac12)M(-\sfrac32)\cdots M(2{-}g)M(1{-}g)\ ,
\end{equation}
which produces a polynomial in the $I_k$ upon repeated use of (\ref{Rdef}) and~(\ref{up}). 
It does not yield a new integral of motion.

In order to understand the relation of $Q(g)$ to the other integrals of motion, 
we compute its square,
\begin{equation} \label{Qsquared}
\begin{aligned}
Q(g)^2 &\= 
M(g{-}1)\cdots M(3{-}g)M(2{-}g)\underbrace{M(1{-}g)M(g{-}1)}M(g{-}2)\cdots M(1{-}g) \\
&\= M(g{-}1)\cdots M(3{-}g)M(2{-}g)\cR(g{-}1) M(g{-}2)M(g{-}3)\cdots M(1{-}g) \\[8pt]
&\= M(g{-}1)\cdots M(3{-}g)\underbrace{M(2{-}g)M(g{-}2)}\cR(g{-}2) M(g{-}3)\cdots M(1{-}g) \\
&\= M(g{-}1)\cdots M(3{-}g)\bigl(\cR(g{-}2)\bigr)^2 M(g{-}3)\cdots M(1{-}g) \\
&\quad\vdots \\
&\= M(g{-}1)M(1{-}g)\bigl(\cR(1{-}g)\bigr)^{2g-2} \= \bigl(\cR(1{-}g)\bigr)^{2g-1} \= 
\bigl(\cR(g)\bigr)^{2g-1}\ ,
\end{aligned}
\end{equation}
and find a polynomial in the $I_k$ integrals again.

For the full enhanced algebra of conserved charges in the case of integer coupling,
we lack the commutators of the new charge $Q$ with the integrals $F_\ell$, 
at any fixed $g\in\NN$,
\begin{align}
\im[Q,F_\ell] \= \sfrac{1}{\ell}\bigl[I_\ell,[C,Q]\bigr] &\=
\sfrac1{\ell}H\bigl[I_\ell,[K,Q]\bigr]+\sfrac1{\ell}\bigl[I_\ell,[K,Q]\bigr]\,H
-\sfrac{1}{2}n(n{-}1)(2g{-}1)Q\,I_\ell \nonumber\\[8pt] \label{qconf}
&\= H\,[\im Q,J_\ell] + [\im Q,J_\ell]\,H
-\sfrac{1}{2}n(n{-}1)(2g{-}1)Q\,I_\ell \ .
\end{align}
Clearly, we need to compute either $[C,Q]$ or $[K,Q]$ or $[J_\ell,Q]$.
The first two do not lead to enlightening expressions,\footnote{
despite the fact that $[\cD_i,K]=x^i$ will simplify $[M(g),K]$} but 
\begin{equation}
\bigl[I_k,[\im Q,J_\ell]\bigr]\=\bigl[\im Q,[I_k,J_\ell]\bigr]\=k\,[Q,I_{k+\ell-2}]\=0
\end{equation}
implies that the third commutator is a polynomial in the Liouville charges and~$Q$.
Being totally antisymmetric under particle exchange, it has to be linear in $Q$, thus
\begin{equation}
\im[Q,J_\ell]\=(2g{-}1)\,Q\,\cG_\ell(I)
\end{equation}
with $\cG_\ell$ being a polynomial in the $I_k$ of conformal weight $\ell{-}2$,
and we have pulled out a convenient factor in its definition such that
the coefficients of this polynomial will not depend on~$g$.
The evaluation of the $\cG_\ell$ can be found in Appendix~\ref{apc}, with the result
\begin{equation}
\cG_\ell\=\sfrac12\sum_{j=0}^{\ell-2}I_{\ell-2-j}I_j - \sfrac{\ell-1}{2}I_{\ell-2}\ ,
\end{equation}
where $I_0=n$.
The lowest instances are
\begin{equation}
\begin{aligned}
&\cG_1=0\ ,\qquad
\cG_2=\sfrac12n(n{-}1)\ ,\qquad
\cG_3=(n{-}1)P\ ,\qquad
\cG_4=(n{-}\sfrac32)2H + \sfrac12 P^2\ ,\\[4pt]
&\cG_5=(n{-}2)I_3 + 2HP\ ,\qquad
\cG_6=(n{-}\sfrac52)I_3 + I_3 P + \sfrac12 4H^2\ .
\end{aligned}
\end{equation}
For the desired commutator it follows that
\begin{equation}
\im[Q,F_\ell] \=(2g{-}1) Q\,\bigl(2\cG_\ell H-\sfrac12n(n{-}1)I_\ell\bigr)
\ =:\ Q\,{\cal C}_\ell(I)\ ,
\end{equation}
consistent with the observation
\begin{equation}
\bigl[I_k,[\im Q,F_\ell]\bigr]\=\bigl[\im Q,[I_k,F_\ell]\bigr]
\=k\,[Q\,,I_{k+\ell-2}I_2-I_k I_\ell] \= 0\ .
\end{equation}
Hence, the first few structure constants read
\begin{equation}
\begin{aligned}
&{\cal C}_1=-(2g{-}1)\sfrac12n(n{-}1)P\ ,\qquad
{\cal C}_2=0\ ,\qquad
{\cal C}_3=(2g{-}1)(n{-}1)\bigl(2HP-\sfrac{n}{2}I_3\bigr)\ ,\\[4pt]
&{\cal C}_4=(2g{-}1)\bigl((4n{-}6)H^2+HP^2-\sfrac12n(n{-}1)I_4\bigr)\ .
\end{aligned}
\end{equation}
It is clear that no further algebraically independent constants of motion can be produced
from commuting the known ones.

In this paper, we discuss the formal commutation properties of operators in 
the quantum Calogero model and do not investigate their proper domains and kernels.
Nevertheless, these are important and interesting issues and relate to the physical
features of the theory. However, they are more naturally studied in the context of
a ${\cal PT}$-symmetric deformation of the Calogero model~\cite{Fring05,Fring12-rev,CorPlyu12}.

Any permutation operator~$s_{ij}$ serves as a suitable $\Z_2$ grading operator for our 
enhanced algebra of conserved charges. Since we only consider observables totally symmetric 
or totally antisymmetric under any particle permutation, only the additional conserved 
charge~$Q$ is odd under this grading,
\begin{equation}
\{ Q,s_{ij}\}=0 \ ,
\end{equation}
producing an ${\cal N}{=}1$ nonlinear superalgebra.\footnote{
An additional supercharge of nonlocal nature can be constructed using as a grading operator 
any two-particle permutation operator. However, the supersymmetric structure does not change radically.}
In contrast with the one-dimensional cases, the anticommutator of the supercharge produces 
a polynomial not only in the Hamiltonian but in {\it all\/} the Liouville integrals. 
The nonlinear supersymmetry algebra has the following form,
\begin{equation}\label{algebra}
\begin{aligned}
& [I_{k},I_{\ell}]=0\ ,\qquad 
\im[I_{k},F_\ell]={\cal A}_{k,\ell}(I)\ ,\qquad 
\im[F_k,F_l]={\cal B}_{k,\ell}(I,F)\ ,\\[4pt] 
& [Q,I_\ell]=0\ ,\qquad 
\im[ Q,F_\ell]=Q\,{\cal C}_\ell(I)\ ,\qquad 
\{Q,Q\}=2\bigl({\cal R}(I)\bigr)^{2g-1}\ ,
\end{aligned}
\end{equation}
where the polynomials are
\begin{equation}\label{allpoly}
\begin{aligned}
{\cal A}_{k,\ell}(I) &\= k\left(I_{k+\ell-2}I_2-I_k I_\ell \right) \\[4pt]
{\cal B}_{k,\ell}(I,F)&\= (k{-}\ell)\sfrac12\{F_{k+\ell-2},I_2\}
-(k{-}2)\sfrac12\{F_k,I_\ell\}+ (\ell{-}2)\sfrac12\{F_\ell,I_k\} \\[4pt]
{\cal C}_\ell(I)&\=(2g{-}1)\bigl({\textstyle\sum_{j=0}^{\ell-2}}I_{\ell-2-j}I_j H
-(\ell{-}1)I_{\ell-2}H-\sfrac12n(n{-}1)I_\ell\bigr)\ .
\end{aligned}
\end{equation}
Such algebras have been identified and applied in various single-particle quantum mechanical systems 
\cite{LeivaPlyush03}--\cite{AndIof12-rev}.

\section{Two particles}

When $n{=}2$, all quantities are easily computed, since after separating the center-of-mass
degree of freedom, one is left with a rank-one system parametrized by the difference variables,
\begin{equation}
x\equiv x^{12}:=x^1{-}x^2\ ,\quad 
2\pa\equiv \pa_{12}:=\pa_1{-}\pa_2 \ ,\quad
\cD_{12}:=\cD_1{-}\cD_2=2\Bigl(\pa-\frac{g}{x}s_{12}\Bigr)\ .
\end{equation}
In terms of these, the conserved charges take the form
\begin{eqnarray}
&& P \= -\im(\pa_1+\pa_2)\ , \\
&& H \= \sfrac14 P^2 - \sfrac14\res\bigl(\cD_{12}^2\bigr)
\= \sfrac14 P^2 - \Bigl(\pa+\frac{g}{x}\Bigr)\Bigl(\pa-\frac{g}{x}\Bigr)
\= \sfrac14 P^2 -\pa^2 + \frac{g(g{-}1)}{x^2}\ ,
\qquad\ph \\[4pt]
&& F_1 \= \sfrac12(x^1{+}x^2)(4H-P^2)+\im(x\pa+\sfrac12)P  \nonumber \\
&& \quad\ \= (x^1\pa_2{-}x^2\pa_1)(\pa_1{-}\pa_2)
+\sfrac12(\pa_1{+}\pa_2)+2g(g{-}1)\frac{x^1{+}x^2}{(x^1{-}x^2)^2}\ ,
\end{eqnarray}
while the $sl(2)$ Casimir reads
\begin{equation}
\begin{aligned}
C &\= \bigl((x^1)^2{+}(x^2)^2\bigr)H+\sfrac12(x^1\pa_1+x^2\pa_2)^2-\sfrac12 \\
&\= -\sfrac12(x^1\pa_2{-}x^2\pa_1)^2-\sfrac12
+g(g{-}1)\frac{(x^1)^2{+}(x^2)^2}{(x^1{-}x^2)^2}\ .
\end{aligned}
\qquad\qquad\qquad\qquad\qquad\ph
\end{equation}
The only nontrivial commutator among the conserved charges is
\begin{equation}
\im[P,F_1]\=4H-P^2\ .
\end{equation}
In this combination, the center of mass is decoupled.

Let us look at the additional charge which appears at integer values of~$g$.
The intertwiner is immediately constructed,
\begin{equation}
M(g)\=2\,\res\bigl(\pa-\sfrac{g}{x}s_{12}\bigr)
\=2\Bigl(\pa-\frac{g}{x}\Bigr)
\=2 x^{g+1}\bigl(\sfrac1{x}\pa\bigr)x^{-g}
\=2 x^g \pa x^{-g}\ .
\end{equation}
To verify the properties of $Q(g)$, it is convenient to remove the center of mass
contributions,
\begin{equation}
I_k(g)\big|_{P=X=0} \ =:\ \tI_k(g) \qquad\Rightarrow\qquad
\tH(g)\= H-\sfrac14P^2 \= -\pa^2+\frac{g(g{-}1)}{x^2}\ ,
\end{equation}
and check that 
\begin{equation}
M(g)\tH(g) \= 2\Bigl(\pa-\frac{g}{x}\Bigr)\Bigl(-\pa^2+\frac{g(g{-}1)}{x^2}\Bigr)
\= 2\Bigl(-\pa^2+\frac{g(g{+}1)}{x^2}\Bigr)\Bigl(\pa-\frac{g}{x}\Bigr) \= \tH(g{+}1)M(g)\ .
\end{equation}
The extra integral of motion is of order $2g{-}1$,
\begin{equation}
Q(g)=M(g{-}1)\cdots M(1{-}g) \qquad\Rightarrow\qquad
Q(g)^2=\bigl(P^2{-}4H(g)\bigr)^{2g-1}=\bigl(-4\tH(g)\bigr)^{2g-1}
\end{equation}
due to \ $M(-g)M(g)=P^2{-}4H(g)$.
By construction, $Q$ commutes with $H$ and $P$, and with $F_1$ it obeys
\begin{equation}
\im[Q,F_1] \= -(2g{-}1)\,Q\,P\ .
\end{equation}
At small values of~$g$, for the full intertwiners defined in~(\ref{Ldef}) one finds
\begin{eqnarray}
L(2) &\!\!=\!\!& 2\bigl(\pa-\sfrac{1}{x}\bigr)\ ,\\ \notag
L(3) &\!\!=\!\!& 4\bigl(\pa^2-\sfrac{3}{x}\pa+\sfrac{3}{x^2}\bigr)\ ,\\ \notag
L(4) &\!\!=\!\!& 8\bigl(\pa^3-\sfrac{6}{x}\pa^2+\sfrac{15}{x^2}\pa-\sfrac{15}{x^3}\bigr)\ ,\\ \notag
L(5) &\!\!=\!\!& 16\bigl(\pa^4-\sfrac{10}{x}\pa^3+\sfrac{45}{x^2}\pa^2-\sfrac{105}{x^3}\pa
+\sfrac{105}{x^4}\bigr)\ ,
\qquad\qquad\qquad\qquad\qquad\qquad\qquad\ph
\end{eqnarray}
and the odd charges take the form
\begin{eqnarray}
Q(1) &\!\!=\!\!& 2\pa\ ,\\ \notag
Q(2) &\!\!=\!\!& 8\bigl(\pa^3-\sfrac{3}{x^2}\pa+\sfrac{3}{x^3}\bigr)\ ,\\ \notag
Q(3) &\!\!=\!\!& 32\bigl(\pa^5-\sfrac{15}{x^2}\pa^3+\sfrac{45}{x^3}\pa^2-\sfrac{45}{x^4}\pa\bigr)\ ,\\ \notag
Q(4) &\!\!=\!\!& 128\bigl(\pa^7-\sfrac{42}{x^2}\pa^5+\sfrac{210}{x^3}\pa^4-\sfrac{315}{x^4}\pa^3
-\sfrac{630}{x^5}\pa^2+\sfrac{2835}{x^6}\pa-\sfrac{2835}{x^7}\bigr)\ ,\\ \notag
Q(5) &\!\!=\!\!& 512\bigl(\pa^9-\sfrac{90}{x^2}\pa^7+\sfrac{630}{x^3}\pa^6-\sfrac{945}{x^4}\pa^5
-\sfrac{9450}{x^5}\pa^4+\sfrac{61425}{x^6}\pa^3-\sfrac{155925}{x^7}\pa^2+\sfrac{155925}{x^8}\pa\bigr)\ .
\end{eqnarray}

\section{Three particles}

For $n{=}3$, one may also remove the center of mass and deal with the remaining rank-two system.
Because the two-dimensional description of the $A_2$ root system lacks the manifest permutation
symmetry however, we keep all three particle coordinates here~\cite{Cal69,CalMar74}.
Abbreviating
\begin{equation}
x^{ij}:=x^i{-}x^j\ ,\quad \pa_{ij}:=\pa_i{-}\pa_j\ ,\quad
\cD_{ij}:=\cD_i{-}\cD_j=\pa_{ij}-\frac{2g}{x^{ij}}s_{ij}+\frac{g}{x^{jk}}s_{jk}+\frac{g}{x^{ki}}s_{ki}
\end{equation}
for $\{i,j,k\}$ being a permutation of $\{1,2,3\}$, the conserved charges read
\begin{eqnarray}
&& P \= -\im\bigl(\pa_1{+}\pa_2{+}\pa_3\bigr)\ , \\[4pt]
&& H \= -\sfrac12(\pa_1^2{+}\pa_2^2{+}\pa_3^2)+g(g{-}1)
\bigl(\sfrac{1}{(x^{12})^2}+\sfrac{1}{(x^{23})^2}+\sfrac{1}{(x^{31})^2}\bigr)\ ,\\
&& I_3 \= \im\bigl(\pa_1^3{+}\pa_2^3{+}\pa_3^3\bigr)-3\im g(g{-}1)\bigl(
\sfrac{\pa_1{+}\pa_2}{(x^{12})^2}+\sfrac{\pa_2{+}\pa_3}{(x^{23})^2}+\sfrac{\pa_3{+}\pa_1}{(x^{31})^2}\bigr)
\ , \qquad\qquad\qquad\qquad\ph\\[4pt]
&& F_1 \= 2(x^1{+}x^2{+}x^3)H + \im(x^1\pa_1{+}x^2\pa_2{+}x^3\pa_3+1)P\ ,\\[4pt]
&& F_3 \= 2J_3 H +\im(x^1\pa_1{+}x^2\pa_2{+}x^3\pa_3+2)I_3
\end{eqnarray}
with
\begin{equation}
J_3 \= -\pa_1x^1\pa_1{-}\pa_2x^2\pa_2{-}\pa_3x^3\pa_3 +g(g{-}1)\bigl(
\sfrac{x^1{+}x^2}{(x^{12})^2}+\sfrac{x^2{+}x^3}{(x^{23})^2}+\sfrac{x^3{+}x^1}{(x^{31})^2}\bigr)\ .
\quad\ \ph
\end{equation}
The $sl(2)$ Casimir takes the form
\begin{equation}
\begin{aligned}
C &\= \bigl((x^1)^2{+}(x^2)^2{+}(x^3)^2\bigr)\,H+\sfrac12
\bigl(x^1\pa_1{+}x^2\pa_2{+}x^3\pa_3+\sfrac32\bigr)
\bigl(x^1\pa_1{+}x^2\pa_2{+}x^3\pa_3-\sfrac12\bigr) \\[6pt]
&\=-\sfrac{1}{2}({\cal J}_{12}^2+{\cal J}_{23}^2+{\cal J}_{31}^2)-\sfrac{3}{8}+
g(g{-}1)\bigl((x^1)^2{+}(x^2)^2{+}(x^3)^2\bigr)
\bigl(\sfrac{1}{(x^{12})^2}{+}\sfrac{1}{(x^{23})^2}{+}\sfrac{1}{(x^{31})^2}\bigr)\ ,
\end{aligned}
\end{equation}
where we used the angular-momentum operators
\begin{equation}
{\cal J}_{ij}\=x^j \pa_i-x^i \pa_j\ .
\end{equation}
The nonvanishing commutators are
\begin{equation}
\begin{aligned}
&\im[P,F_1]\=6H-P^2\ ,\quad \im[I_3,F_1]\=12H^2-3I_3 P\ ,\quad \im[P,F_3]\=4H^2-I_3 P\ ,\\
&\im[I_3,F_3]\=-3I_3^2+8I_3HP+12H^3-12H^2P^2+HP^4\ ,\\
&\im[F_1,F_3]\=\sfrac12(F_1I_3+I_3F_1+F_3P+PF_3)\ .
\end{aligned}
\end{equation}

Applying the formula~(\ref{up}), we find the intertwiner
\begin{align}
M(g) &\= \res\bigl(\cD_{12}(g)\cD_{23}(g)\cD_{31}(g)\bigr) \\[6pt] \nonumber 
&\= \Delta^g \bigl(
\pa_{12}\pa_{23}\pa_{31}+\sfrac{g}{x^{12}}\pa_{12}^2+\sfrac{g}{x^{23}}\pa_{23}^2+\sfrac{g}{x^{31}}\pa_{31}^2 
-\sfrac{2g}{(x^{12})^2}\pa_{12} -\sfrac{2g}{(x^{23})^2}\pa_{23} -\sfrac{2g}{(x^{31})^2}\pa_{31}
\bigr)\,\Delta^{-g}
\end{align}
with $\Delta=x^{12}x^{23}x^{31}$.
Its explicit form is
\begin{equation}\label{mg3}
\begin{aligned}
M(g)&\=\pa_{12}\pa_{23}\pa_{31}
-\sfrac{2g}{x^{12}}\pa_{23}\pa_{31}
-\sfrac{2g}{x^{23}}\pa_{31}\pa_{12}
-\sfrac{2g}{x^{31}}\pa_{12}\pa_{23} \\[4pt] & \
+\sfrac{4g^2}{x^{12}x^{23}}\pa_{31}
+\sfrac{4g^2}{x^{23}x^{31}}\pa_{12}
+\sfrac{4g^2}{x^{31}x^{12}}\pa_{23}
-\sfrac{g(g{-}1)}{(x^{12})^2}\pa_{12}
-\sfrac{g(g{-}1)}{(x^{23})^2}\pa_{23}
-\sfrac{g(g{-}1)}{(x^{31})^2}\pa_{31} \\[4pt] & \
-\sfrac{6\,g^2(g{+}1)}{x^{12}x^{23}x^{31}}
+2g(g{-}1)(g{+}2)\Bigl(\sfrac{1}{(x^{12})^3}+
\sfrac{1}{(x^{23})^3}+\sfrac{1}{(x^{31})^3}\Bigr)\ .
\end{aligned}
\end{equation}
For $g{=}1$ this reduces to eq.~(19) of~\cite{ChaVes90} (after correcting a typo there).
$M(g)$ is a third-order partial differential operator and satisfies the relation
\begin{equation}
M(-g)M(g) \= 
3I_3^2-12I_3HP+\sfrac43I_3P^3-4H^3+14H^2P^2-3HP^4+\sfrac16P^6\ .
\end{equation}
The additional conserved charge is of order $3(2g{-}1)$,
\begin{equation}
Q(g)=M(g{-}1)\cdots M(1{-}g) \qquad\Rightarrow\qquad
Q(g)^2=\bigl(M(-g)M(g)\bigr)^{2g-1}\ .
\end{equation}
Its nontrivial commutation relations are
\begin{equation}
\begin{aligned}
\im[Q,F_1]&\=-3(2g{-}1)\,Q\,P\ , \\
\im[Q,F_3]&\=-3(2g{-}1)\,Q\bigl(I_3-\sfrac{4}{3}H P\bigr).
\end{aligned}
\end{equation}
The lowest full intertwiner reads
\begin{equation}
\begin{aligned}
L(2)\=M(1)&\=\pa_{12}\pa_{23}\pa_{31}
-\sfrac{2}{x^{12}}\pa_{23}\pa_{31}
-\sfrac{2}{x^{23}}\pa_{31}\pa_{12}
-\sfrac{2}{x^{31}}\pa_{12}\pa_{23} \\[4pt] & \quad
+\sfrac{4}{x^{12}x^{23}}\pa_{31}
+\sfrac{4}{x^{23}x^{31}}\pa_{12}
+\sfrac{4}{x^{31}x^{12}}\pa_{23}
-\sfrac{12}{x^{12}x^{23}x^{31}}\ ,
\end{aligned}
\end{equation}
and the first two odd charges take the form
\begin{align}
Q(1) &\= \pa_{12}\pa_{23}\pa_{31}\ ,\\[8pt]
Q(2) &\=\sfrac{1}{6} \pa_{12}^3\pa_{23}^3\pa_{31}^3 -\sfrac{3}{(x^{12})^2}\bigl(
\pa_{12}^3\pa_{23}^2\pa_{31}^2{+}2 \pa_{12}\pa_{23}^3\pa_{31}^3 \bigr) \notag
+\sfrac{12}{(x^{12})^3}\bigl(\pa_{12}^2\pa_{23}^3 \pa_{31}{+}\pa_{23}^3\pa_{31}^3
{+} 4 \pa_{12}^2 \pa_{23}^2 \pa_{31}^2 \bigr) \\ \notag
& +12\bigl( -\sfrac{1}{(x^{12})^4} +\sfrac{2}{(x^{12})^2(x^{31})^2} \bigr)\pa_{12}^3\pa_{23}^2
+12 \bigl( \sfrac{22}{(x^{23})^4} - \sfrac{15}{(x^{12})^4}  - \sfrac{14}{(x^{12})^2(x^{23})^2} \bigr)
\pa_{12}\pa_{23}^2 \pa_{31}^2 \\ \notag
& -720 \bigl( \sfrac{2}{(x^{12})^5} -\sfrac{1}{(x^{12})^3(x^{31})^2} -\sfrac{1}{(x^{12})^2(x^{31})^3}\bigr)
\pa_{12}^3\pa_{23}  \\ \notag
& -360  \bigl( \sfrac{3}{(x^{12})^5}-\sfrac{1}{(x^{31})^5}  -\sfrac{1}{(x^{12})^3(x^{31})^2}  
-\sfrac{3}{(x^{12})^2(x^{31})^3} \bigr) \pa_{12}^2\pa_{23}^2 
+\sfrac{4320}{(x^{12})^2(x^{31})^4}\pa_{12}^2\pa_{23} \\ \notag
& -120 \bigl( \sfrac{35}{(x^{12})^6} +\sfrac{28}{(x^{23})^6} -\sfrac{16}{(x^{12})^3(x^{23})^3}  
+\sfrac{10}{(x^{23})^3(x^{31})^3} +\sfrac{24}{(x^{12})^2(x^{31})^4}  \bigr) \pa_{12}^3
+\sfrac{5760}{(x^{12})^2(x^{31})^4}\pa_{12}\pa_{23} \pa_{31}  \\ \notag
& +720 \bigl( \sfrac{35}{(x^{12})^7}-\sfrac{14}{(x^{23})^7} -\sfrac{10}{(x^{12})^5(x^{23})^2} 
-\sfrac{8}{(x^{12})^4(x^{23})^3}+\sfrac{14}{(x^{12})^3(x^{23})^4} -\sfrac{2}{(x^{12})^2(x^{23})^5}\bigr)
\pa_{12}^2\\ \notag
& -4320 \bigl( \sfrac{21}{(x^{12})^8} +\sfrac{46}{(x^{12})^7(x^{23})} -\sfrac{30}{(x^{12})^6(x^{23})^2} 
+\sfrac{8}{(x^{12})^5(x^{23})^3} -\sfrac{4}{(x^{12})^3(x^{23})^5}  \bigr)\pa_{12} \\
&+\sfrac{181440}{(x^{12})^9} +\sfrac{60480}{(x^{12})^7x^{23}x^{31}}
\ + \ \textrm{ all permutations in $(123)$} \ .
\end{align}

\section{Four particles}

This is the simplest case which cannot be separated into one-dimensional systems.
With the same abbreviations as in the previous section but 
\begin{equation}
\cD_{12}\=\pa_{12}-\frac{2g}{x^{12}}s_{12}-\frac{g}{x^{13}}s_{13}+\frac{g}{x^{23}}s_{23}
-\frac{g}{x^{14}}s_{14}+\frac{g}{x^{24}}s_{24} \qquad\textrm{and permutations}\ ,
\end{equation}
we have the following integrals of motion,
\begin{align}
P &\= -\im\bigl(\pa_1{+}\pa_2{+}\pa_3{+}\pa_4\bigr)\ , \\[4pt]
H &\= -\sfrac12(\pa_1^2{+}\pa_2^2{+}\pa_3^2{+}\pa_4^2)+g(g{-}1)
\bigl(\sfrac{1}{(x^{12})^2}+\sfrac{1}{(x^{13})^2}+\sfrac{1}{(x^{14})^2}
+\sfrac{1}{(x^{23})^2}+\sfrac{1}{(x^{24})^2}+\sfrac{1}{(x^{34})^2}\bigr)\ ,\\
I_3 &\= \im\bigl(\pa_1^3{+}\pa_2^3{+}\pa_3^3{+}\pa_4^3\bigr)-3\im g(g{-}1)\bigl(
\sfrac{\pa_1{+}\pa_2}{(x^{12})^2}+\sfrac{\pa_1{+}\pa_3}{(x^{13})^2}+\sfrac{\pa_1{+}\pa_4}{(x^{14})^2}+
\sfrac{\pa_2{+}\pa_3}{(x^{23})^2}+\sfrac{\pa_2{+}\pa_4}{(x^{24})^2}+\sfrac{\pa_3{+}\pa_4}{(x^{34})^2}\bigr)
\ ,\\ \notag
I_4 &\= \pa_1^4{+}\pa_2^4{+}\pa_3^4{+}\pa_4^4 \\ \notag
&-4g(g{-}1) \bigl(
\sfrac{\pa_1^2{+}\pa_2^2{+}\pa_1\pa_2}{(x^{12})^2}{+}
\sfrac{\pa_1^2{+}\pa_3^2{+}\pa_1\pa_3}{(x^{13})^2}{+}
\sfrac{\pa_1^2{+}\pa_4^2{+}\pa_1\pa_4}{(x^{14})^2}{+}
\sfrac{\pa_2^2{+}\pa_3^2{+}\pa_2\pa_3}{(x^{23})^2}{+}
\sfrac{\pa_2^2{+}\pa_4^2{+}\pa_2\pa_4}{(x^{24})^2}{+}
\sfrac{\pa_3^2{+}\pa_4^2{+}\pa_3\pa_4}{(x^{34})^2} \bigr) \\
&+4g(g{-}1)\bigl(
\sfrac{\partial_1-\partial_2}{(x^{12})^3}+\sfrac{\partial_3-\partial_1}{(x^{31})^3}+
\sfrac{\partial_1-\partial_4}{(x^{14})^3}+\sfrac{\partial_2-\partial_3}{(x^{23})^3}+
\sfrac{\partial_2-\partial_4}{(x^{24})^3}+\sfrac{\partial_3-\partial_4}{(x^{34})^3}\bigr) \\ \notag
&+2(g{+}2)g(g{-}1)(g{-}3)\bigl(
\sfrac1{(x^{12})^4}+\sfrac1{(x^{31})^4}+\sfrac1{(x^{14})^4}+
\sfrac1{(x^{23})^4}+\sfrac1{(x^{24})^4}+\sfrac1{(x^{34})^4} \bigr)\\ \notag
&+4g^2(g{-}1)^2\bigl(
\sfrac1{(x^{12})^2(x^{31})^2}+\sfrac1{(x^{12})^2(x^{14})^2}+\sfrac1{(x^{31})^2(x^{14})^2}+
\sfrac1{(x^{21})^2(x^{23})^2}+\sfrac1{(x^{21})^2(x^{24})^2}+\sfrac1{(x^{23})^2(x^{24})^2} \\ \notag
&\qquad\qquad\quad\! +
\sfrac1{(x^{31})^2(x^{32})^2}+\sfrac1{(x^{31})^2(x^{34})^2}+\sfrac1{(x^{32})^2(x^{34})^2}+
\sfrac1{(x^{41})^2(x^{42})^2}+\sfrac1{(x^{41})^2(x^{43})^2}+\sfrac1{(x^{42})^2(x^{43})^2}\bigr)\ ,\\[4pt]
F_1&\=2(x^1{+}x^2{+}x^3{+}x^4)H+\im(x^1\pa_1{+}x^2\pa_2{+}x^3\pa_3{+}x^4\pa_4+\sfrac32)P\ ,\\[4pt]
F_3&\=2J_3 H +\im(x^1\pa_1{+}x^2\pa_2{+}x^3\pa_3{+}x^4\pa_4+\sfrac52)I_3\ ,\\[4pt]
F_4&\=2J_4 H +\im(x^1\pa_1{+}x^2\pa_2{+}x^3\pa_3{+}x^4\pa_4+3)I_4\ , 
\end{align}
where derivatives in fraction numerators are understood to be ordered to the right, and with
\begin{align}
J_3 &\= -\,\pa_1x^1\pa_1-\pa_2x^2\pa_2-\pa_3x^3\pa_3-\pa_4x^4\pa_4 \\ \notag
&\qquad + g(g{-}1)\bigl(
\sfrac{x^1{+}x^2}{(x^{12})^2}+\sfrac{x^1{+}x^3}{(x^{13})^2}+\sfrac{x^1{+}x^4}{(x^{14})^2}+
\sfrac{x^2{+}x^3}{(x^{23})^2}+\sfrac{x^2{+}x^4}{(x^{24})^2}+\sfrac{x^3{+}x^4}{(x^{34})^2}\bigr)\ ,\\[4pt]
J_4 &\= \im(x^1\pa_1^3+x^2\pa_2^3+x^3\pa_3^3+x^4\pa_4^3)
+\sfrac32\im(\pa_1^2+\pa_2^2+\pa_3^2+\pa_4^2) \\ \notag
&\qquad-\im g(g{-}1)\bigl(
\sfrac{(2x^1+x^2)\pa_1+(2x^2+x^1)\pa_2+1}{(x^{12})^2}{+}
\sfrac{(2x^1+x^3)\pa_1+(2x^3+x^1)\pa_3+1}{(x^{13})^2}{+}
\sfrac{(2x^1+x^4)\pa_1+(2x^4+x^1)\pa_4+1}{(x^{14})^2} \\ \notag
&\qquad\qquad\qquad +
\sfrac{(2x^2+x^3)\pa_2+(2x^3+x^2)\pa_3+1}{(x^{23})^2}{+}
\sfrac{(2x^2+x^4)\pa_2+(2x^4+x^2)\pa_4+1}{(x^{24})^2}{+}
\sfrac{(2x^3+x^4)\pa_3+(2x^4+x^3)\pa_4+1}{(x^{34})^2}\bigr)\ .
\end{align}

The $sl(2)$ Casimir reads
\begin{align}
C&=\bigl((x^1)^2{+}(x^2)^2{+}(x^3)^2{+}(x^4)^2\bigr) H +
\sfrac12\bigl(x^1\pa_1{+}x^2\pa_2{+}x^3\pa_3{+}x^4\pa_4{+}2\bigr)
\bigl(x^1\pa_1{+}x^2\pa_2{+}x^3\pa_3{+}x^4\pa_4\bigr) \nonumber \\
&=-\sfrac{1}{2}\sum_{i<j}^4{\cal J}_{ij}^2+g(g{-}1)\sum_{k=1}^4(x^k)^2
\sum_{i<j}^4\frac{1}{(x^{ij})^2}\ .
\end{align}

\begin{figure}[ht!]
\centering
\includegraphics[scale=1]{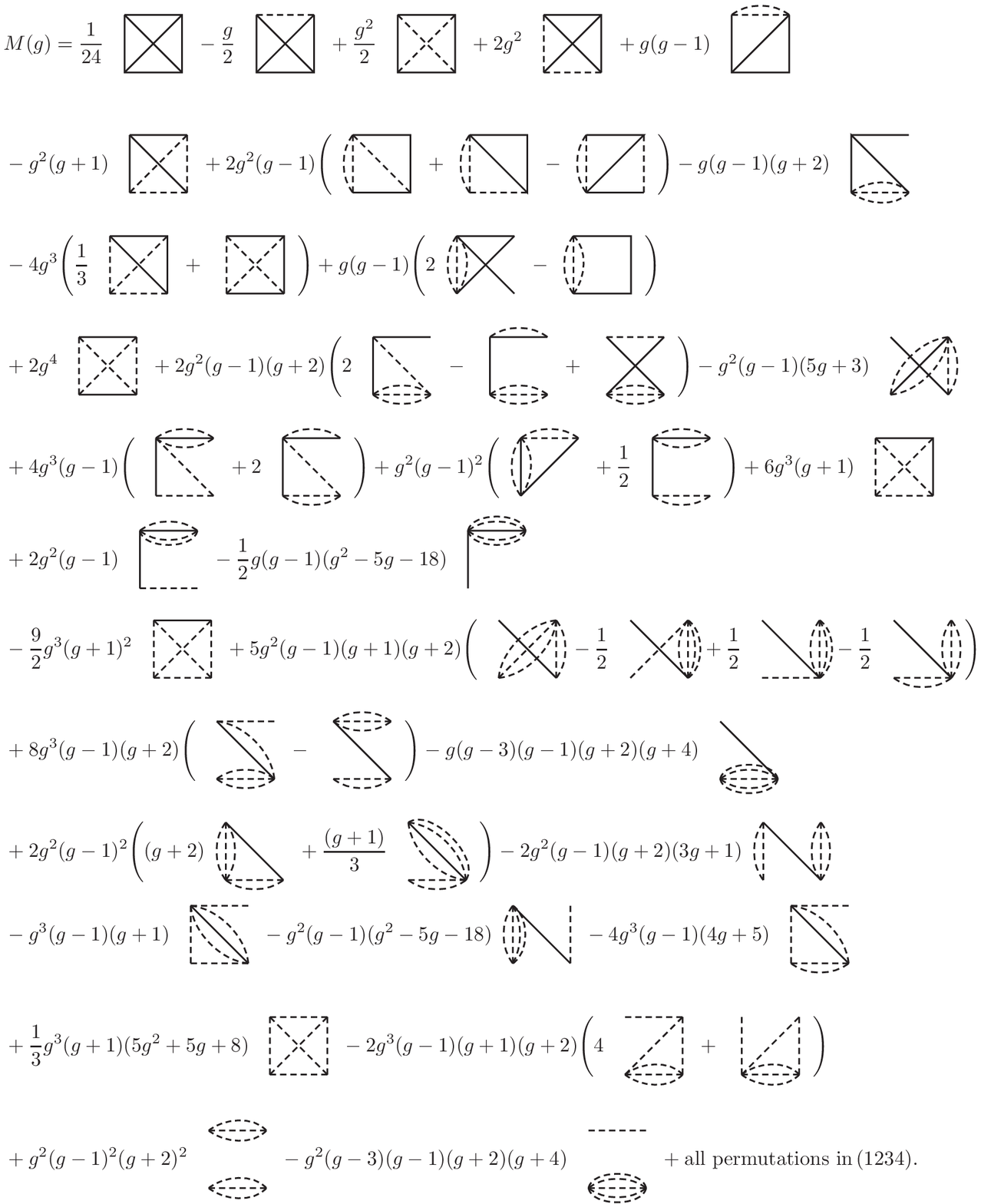}
\caption{$M(g)$ for the four-particle case.}
\label{mg4aa}
\end{figure}

The list of nonvanishing commutators is the following,
\begin{align}
\im[P,F_1]&=8H-P^2\ ,\quad \im[I_3,F_1]=12H^2-3I_3 P\ ,\quad \im[P,F_3]=4H^2-I_3 P\ ,\notag\\
\im[I_3,F_3]&=6I_4 H-3I_3^2\ ,\quad \im[P,F_4]=2I_3H-I_4P\ ,\quad \im[I_4,F_1]=8I_3H-4I_4P\ ,\notag\\
\im[I_3,F_4]&=-3I_4I_3+\sfrac{15}2I_4HP+10I_3H^2-5I_3HP^2-15H^3P+5H^2P^3-\sfrac14HP^5\ ,\notag\\
\im[I_4,F_3]&=-4I_4I_3+10I_4HP+\sfrac{40}3I_3H^2-\sfrac{20}3I_3HP^2-20H^3P
+\sfrac{20}3H^2P^3-\sfrac13HP^5\ ,\notag\\
\im[I_4,F_4]&=-4I_4^2{+}12I_4H^2{+}6I_4HP^2{+}\sfrac83I_3^2H{-}\sfrac{16}3I_3HP^3{-}8H^4
{-}12H^3P^2{+}6H^2P^4{-}\sfrac13HP^6\ ,\notag\\
\im[F_1,F_3]&=\sfrac12(F_1I_3+I_3F_1+F_3P+PF_3)\ ,\\
\im[F_1,F_4]&=\sfrac12(F_1I_4+I_4F_1+2F_4P+2PF_4-12F_3H)\ ,\notag\\
\im[F_3,F_4]&=F_4I_3+I_3F_4-2F_4HP-\sfrac12F_3I_4-\sfrac12I_4F_3-
F_3H(2H{-}P^2)-\sfrac12I_4HF_1+H^3F_1\notag\\
&\ \ +\sfrac13I_3H(F_1P{+}PF_1)-\sfrac13H^2(F_1P^2{+}PF_1P{+}P^2F_1)
+\sfrac1{60}H(F_1P^4{+}\ldots{+}P^4F_1)\ .\notag
\end{align}
The intertwiner is now a differential operator of order six,
\begin{equation} \label{M4}
\begin{aligned}
M(g) &\= \res\bigl(\cD_{12}(g)\cD_{31}(g)\cD_{14}(g)\cD_{23}(g)\cD_{24}(g)\cD_{34}(g)\bigr)
\end{aligned}
\end{equation}
whose explicit form is represented below using a graphical short-hand notation in 
Figure~\ref{mg4aa}. 
In each diagram, the four particle labels $i=1,2,3,4$ are carried by the corners of a square.
A solid line connecting corner~$i$ with corner~$j$ represents a derivative~$\partial_{ij}$,
and a dashed line between corner~$i$ and corner~$j$ stands for a pole~$\sfrac{1}{x_{ij}}$.
The sign ambiguity is fixed by taking a cyclic ordering for the labels $i,j\in\{1,2,3\}$ and
by always keeping the label~$4$ in the second slot. In other words, we put $i<j$ except
for the pair~$(3,1)$. To illustrate this notation by an example, we translate the third diagram 
in the expression below,

\begin{figure}[ht!]
\centering
\includegraphics[scale=1]{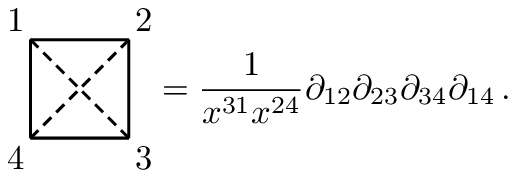}
\end{figure}

We have checked that the expression~(\ref{M4}) intertwines with the Hamiltonian and that it squares to
\begin{equation} \label{12order}
\begin{aligned}
M(\!&{-}g)M(g)\=\\ \sfrac1{576}\bigl( &
P^{12}-48P^{10}H+840P^8H^2-6368P^6H^3+19344P^4H^4-21888P^2H^5+4608H^6\\
+\ & 40P^9I_3-1296P^7HI_3+12576P^5H^2I_3-33344P^3H^3I_3+24576PH^4I_3\\
+\ & 544P^6I_3^2-9024P^4HI_3^2+16896P^2H^2I_3^2-4352H^3I_3^2+2496P^3I_3^3\\
-\ & 1152PHI_3^3-192I_3^4-36P^8I_4+1152P^6HI_4-10800P^4H^2I_4+24768P^2H^3I_4\\
-\ & 11520H^4I_4-1008P^5I_3I_4+16416P^3HI_3I_4-25344PH^2I_3I_4-7200P^2I_3^2I_4\\
+\ & 2304HI_3^2I_4+468P^4I_4^2-7488P^2HI_4^2+9216H^2I_4^2+6912PI_3I_4^2-2304I_4^3
\bigr)\ .
\end{aligned}
\end{equation}
The additional conserved charge $Q(g)$ is a differential operator of order $6(2g{-}1)$, which equals 18
in the simplest nontrivial case of~$g{=}2$. 
It squares to the $(2g{-}1)$th power of the 12th-order polynomial~(\ref{12order}) and obeys
\begin{equation}
\begin{aligned}
\im[Q,F_1]&=-6(2g{-}1)\,Q\,P\ ,\\
\im[Q,F_3]&=-6(2g{-}1)\,Q\,\bigl(I_3-HP\bigr)\ ,\\
\im[Q,F_4]&=-6(2g{-}1)\,Q\,\bigl(I_4-\sfrac{5}{3}H^2-\sfrac{1}{6}HP^2\bigr)\ .
\end{aligned}
\end{equation}

We remark that the isomorphism between the $A_3$ and $D_3$ Lie algebras can be employed to pass
to another distinguished coordinate basis \cite{Dunkl08,KriLePo08,KriLe10},
\begin{equation} 
y^\alpha \= M^\alpha_{\ \ i}\,x^i \qquad\textrm{with}\qquad
\bigl(M^\alpha_{\ \ i}\bigr) \= \sfrac12\,\Bigl(\begin{smallmatrix}
1 & \ph1 &   -1 &   -1 \\
1 &   -1 & \ph1 &   -1 \\
1 &   -1 &   -1 & \ph1 
\end{smallmatrix}\Bigr)
\qquad\textrm{and}\qquad \alpha=1,2,3\ ,
\end{equation}
which manifestly decouples the center of mass defined in (\ref{calmom})
but destroys the permutation symmetry of the coordinate labels.
It remains to be seen whether the coordinates $y^\alpha$ allow for a simplification 
of the four-particle expressions above.

\section{Outlook}

The existence of the additional integral of motion $Q$ for integer values of $g$ in the rational Calogero 
model and its analogy with single-particle models suggest several directions for further investigation. 

Beyond the rational model, there exist the well-known trigonometric, hyperbolic and elliptic generalizations
of the inverse-square interaction. All these models also display Liouville integrals of motion and an 
intertwining operator like~$M(g)$. It is natural to  investigate the existence and nature of the operator~$Q$
as well as the corresponding modification of the supersymmetry structure presented here. 
Although these models are not superintegrable, so integrals like $F_\ell$ will be absent, they are all related
with finite-gap potentials for special values of the coupling parameters. In single-particle finite-gap systems
(or the two-particle case discussed here for integer values of~$g$), the conserved quantity analogous to~$Q$
is the Lax operator. The properties of this Lax operator (for instance its kernel) depend on the type of potential
being considered. Hence, we expect differences in the supersymmetry structure for the distinct interaction
potentials in the quantum integrable many-body models as well. 

In the same spirit, the spectrum of single-particle finite-gap systems and the nature of the Lax operator
change under a ${\cal PT}$ deformation~\cite{CorPlyu12}. Such a generalization deserves to be explored
also in multi-particle integrable systems with a supersymmetry charge~$Q$~\cite{Fring05,Fring12-rev}.

Finally, the analysis of possible factorizations of the operator~$Q$ is mathematically interesting.  
In the single-particle case, the Lax operator naturally induces a factorization which reveals the existence 
of two additional nonlinear supercharges when a matrix Hamiltonian system is considered~\cite{CJP}. 
One can try to generalize this situation to the Calogero model and its cousins.

\bigskip

\section*{Acknowledgments}
\nopagebreak

\noindent
We acknowledge useful discussions with Charles Dunkl, Misha Feigin and Sourya Ray.
This work was partially supported by
the Volkswagen Foundation under grant I/84~496,
by the Deutsche Forschungsgemeinschaft under grant LE 838/12-1,
by the Fondecyt grants 1130017, 11121651, by the Conicyt grant  79112034 and by DICYT (USACH). 
O.L. is grateful for warm hospitality at CECs in Valdivia and
at USACH in Santiago, where this work was begun during a sabbatical leave.
F.C. and M.P.  wish to thank for kind hospitality at Leibniz Universit\"at Hannover,
which was also extended to F.C. by USACH in Santiago and to M.P. by CECs in Valdivia. 
The Centro de Estudios Cient\'{\i}ficos (CECs) is funded by the Chilean government
through the Centers of Excellence Base Financing Program of Conicyt.

\bigskip

\begin{appendix}

\section{Appendix: formulae in the `potential-free frame'}

It is sometimes convenient to perform a similarity transformation of an operator $A$ by the
$g$th power of the Vandermonde determinant,
\begin{equation} \label{similarity}
\widehat{A}\ :=\ \De^{-g} A\,\De^g \qquad\textrm{with}\qquad
\De \= \prod_{i<j} (x^i-x^j)\ .
\end{equation}
The transformed Dunkl operators take the simple form
\begin{equation}
\widehat{\pi}_i \= p_i + \im\sum_{j(\neq i)} \frac{g}{x^i{-}x^j}(s_{ij}{-}1)
\qquad\Leftrightarrow\qquad
\widehat{\cD}_i \= \pa_i + \sum_{j(\neq i)} \frac{g}{x^i{-}x^j}(1{-}s_{ij})\ ,
\end{equation}
and the transformed Hamiltonian looses its potential term, 
acquiring instead a first-order derivative term,
\begin{equation}\label{hhat}
\widehat{H}\=-\sfrac12\res\Bigl(\sum_i \widehat{\cD}_i^2\Bigr) 
\= -\sfrac12\sum_i\pa_i^2 - \sum_{i<j}\frac{g}{x^i{-}x^j}(\pa_i{-}\pa_j)\ .
\end{equation}
For this reason it is called the `potential-free frame'.
Note, however, that the `duality' symmetry between $g$ and $1{-}g$ is hidden in this frame.

Heckman's intertwining formulae~\cite{Heckman,Opdam} were in fact derived in this frame,
\begin{eqnarray}\label{heckop}
\hM(g)\,\hI_k(g) \= \hI_k(g{+}1)\,\hM(g) &&\quad\textrm{for}\qquad
\hM(g)\=\res\Bigl(\De^{-1}\prod_{i<j}(\hcD_i{-}\hcD_j)(g)\Bigr)\ , \\
\hM(g)^*\,\hI_k(g{+}1) \= \hI_k(g)\,\hM(g)^* &&\quad\textrm{for}\qquad
\hM(g)^*\=\res\Bigl(\prod_{i<j}(\hcD_i{-}\hcD_j)(g)\;\De\Bigr)\ .
\end{eqnarray}
Since $\hM(g)$ is an intertwiner, the definition (\ref{similarity}) must be
modified to $M(g)=\De^{g+1}\hM(g)\De^{-g}$ here.
The intertwiner $\hM(g)$ is totally symmetric in the $x^i$ variables and
a differential operator of order $\sfrac12n(n{-}1)$.

\section{Appendix: dependent Liouville charges}\label{apb}

The dependent integrals $I_{n+1}$ can be expressed in terms of the independent ones by means of
\begin{equation} \label{dependent}
I_k\=\tr\left( \begin{matrix}
\ph e_1 & 1 & 0 & \cdots & 0 \\
-e_2    & 0 & 1 & \cdots & 0 \\
\vdots & \vdots & \vdots & \ddots & \vdots \\
\pm e_{n-1} & 0 & 0 & \cdots & 1 \\
\mp e_n & 0 & 0 & \cdots & 0 
\end{matrix} \right)^{\,\textstyle k} \quad\textrm{with}\quad
e_\ell \= \frac1{\ell!} \det \left( \begin{matrix}
I_1 & 1 & 0 & \cdots & 0 \\
I_2 & I_1 & 2 & \cdots & 0 \\
\vdots & \vdots & \vdots & \ddots & \vdots \\
I_{\ell-1} & I_{\ell-2} & I_{\ell-3} & \cdots & \ell{-}1 \\
I_\ell & I_{\ell-1} & I_{\ell-2} & \cdots & I_1 
\end{matrix} \right) \ ,
\end{equation}
where the sign choice correlates with $n$ being even or odd.
In this way, one finds that
\begin{equation}
\begin{aligned}
n=2: \ 
I_3&=3HP-\sfrac12P^3\ ,\quad
I_4=2H^2+2HP^2-\sfrac12P^4\ ,\quad
I_5=5H^2P-\sfrac14P^5\ ,\\
I_6&=2H^3+6H^2P^2-\sfrac32HP^4\ ,\quad
I_7=7H^3P+\sfrac72H^2P^3-\sfrac74HP^5+\sfrac18P^7\ ,\\
F_3&=HF_1-\sfrac16(F_1P^2{+}PF_1P{+}P^2F_1)\ ,\\
F_4&= \sfrac12H(F_1P{+}PF_1)-\sfrac18(F_1P^3{+}P^2F_1P+PF_1P^2{+}P^3F_1)\ ,\\
F_5&=H^2F_1-\sfrac1{20}(F_1P^4{+}PF_1P^3{+}P^2F_1P^2{+}P^3F_1P{+}P^4F_1)\ ,
\ldots\ ,\\[4pt]
n=3: \ 
I_4&=\sfrac43 I_3 P + 2H^2-2H P^2+\sfrac16 P^4\ ,\quad
I_5 = \sfrac53 I_3 H + \sfrac56 I_3 P^2 - \sfrac53 H P^3 +\sfrac16 P^5\ ,\\
I_6&=\sfrac13 I_3^2+2I_3HP+\sfrac13 I_3P^3+2H^3-3H^2P^2-\sfrac12 HP^4+\sfrac1{12}P^6\ ,\\
F_4&=F_3 P+\sfrac13 I_3 F_1-\sfrac12H(F_1P{+}PF_1)+\sfrac1{24}(F_1P^3{+}PF_1P^2{+}P^2F_1P{+}P^3F_1)\ ,\\
F_5&=\sfrac12F_3(2H{+}P^2)+\sfrac16I_3(F_1P{+}PF_1)-\sfrac13H(F_1P^2{+}PF_1P{+}P^2F_1) \\
& \ \ +\sfrac1{30}(F_1P^4{+}PF_1P^3{+}P^2F_1P^2{+}P^3F_1P{+}P^4F_1)\ ,
\ldots\ ,\\[4pt]
n=4: \
I_5&=\sfrac54 I_4P+\sfrac53 I_3H-\sfrac56 I_3P^2-\sfrac52 H^2P+\sfrac56 HP^3-\sfrac1{24}P^5\ ,\\
I_6&=\sfrac32 I_4H+\sfrac34 I_4P^2+\sfrac13 I_3^2-\sfrac23 I_3P^3-H^3-\sfrac32 H^2P^2
+\sfrac34 HP^4-\sfrac1{24}P^6\ ,\\
F_5&= F_4P+\sfrac12F_3(2H{-}P^2)+\sfrac14I_4F_1-\sfrac16I_3(F_1P{+}PF_1)-\sfrac12H^2F_1\\
& \ \ +\sfrac16H(F_1P^2{+}PF_1P{+}P^2F_1)-\sfrac1{120}(F_1P^4{+}\ldots{+}P^4F_1)\ ,\ldots\ ,
\end{aligned}
\end{equation}
and so on.

\section{Appendix: derivation of the polynomials $\cG_\ell(I)$}\label{apc}

The task of this appendix is the computation of the polynomials $\cG_\ell(I)$ appearing in
\begin{equation}
\im\bigl[Q(g),J_\ell(g)\bigr]\=(2g{-}1)\,Q(g)\,\cG_\ell\bigl(I(g)\bigr)\ .
\end{equation}
The factor of $2g{-}1$ stems from the fact that $Q(g)$ is a product of $2g{-}1$
intertwiners $M(g{-}j)$ for $j=1,\ldots,2g{-}1$. 
The remaining dependence on $g$ can only hide in $Q(g)$ and $I(g)$. 
To find the polynomials $\cG_\ell(I)$, it therefore suffices to investigate the free case, 
$g=1$, i.e.
\begin{equation}
\im\bigl[Q(1),J_\ell(1)\bigr]\=Q(1)\,\cG_\ell\bigl(I(1)\bigr)\ ,
\end{equation}
with
\begin{equation}
Q(1)\ \sim\ \prod_{i<j}(p_i{-}p_j) \qquad\textrm{and}\qquad
J_\ell(1)\=\sum_i\bigl(x^i p_i^{\ell-1}\bigr)_{\textrm{sym}}\ ,
\end{equation}
where the label `sym' denotes symmetric or Weyl ordering, with weight $1/\ell$.
In the commutator, each term in the symmetrized sum $J_\ell(1)$ contributes
identically. Hence, the ordering does not matter, and one finds
\begin{equation}
\im\Bigl[ {\textstyle\prod_{i<j}}(p_i{-}p_j)\,,\,
{\textstyle\sum_k}(x^k p_k^{\ell-1})_{\textrm{sym}}\Bigr] \=
\Bigl({\textstyle\prod_{i<j}}(p_i{-}p_j)\Bigr)\,\sum_{k\neq m}\frac{p_k^{\ell-1}}{p_k{-}p_m}\ .
\end{equation}
We read off that
\begin{equation}
\begin{aligned}
\cG_\ell &\= \sfrac12\sum_{k\neq m}\frac{p_k^{\ell-1}{-}p_m^{\ell-1}}{p_k-p_m}
\=\sfrac12\sum_{k\neq m}\sum_{j=0}^{\ell-2} p_k^{\ell-2-j}p_m^j
\=\sfrac12\sum_{k,m}\sum_{j=0}^{\ell-2} p_k^{\ell-2-j}p_m^j 
- \sfrac{\ell{-}1}{2}\sum_k p_k^{\ell-2} \\[4pt]
&\=\sfrac12\sum_{j=0}^{\ell-2}\bigl({\textstyle\sum_k}p_k^{\ell-2-j}\bigr)
\bigl({\textstyle\sum_m} p_m^j\bigr) - \sfrac{\ell{-}1}{2}{\textstyle\sum_k} p_k^{\ell-2}
\=\sfrac12\sum_{j=0}^{\ell-2} I_{\ell-2-j}I_j - \sfrac{\ell{-}1}{2} I_{\ell-2}\ ,
\end{aligned}
\end{equation}
as claimed. It is obvious that
\begin{equation}
\cG_1=0\ ,\qquad \cG_2=\sfrac12\sum_{k\neq m}1=\sfrac12n(n{-}1)\ ,\qquad
\cG_3=\sfrac12\sum_{k\neq m}(p_k+p_m)=(n{-}1)P\ .
\end{equation}
For $\ell>2$, we may split off the first and last term in the sum and insert $I_0=n$,
\begin{equation}
\cG_\ell\=\bigl(n{-}\sfrac{\ell-1}{2}\bigr)I_{\ell-2}
+I_{\ell-3}I_1+I_{\ell-4}I_2+I_{\ell-5}I_3+\ldots+
\begin{cases} 
I_{\lceil(\ell-2)/2\rceil}I_{\lfloor(\ell-2)/2\rfloor} & \textrm{for $\ell$ odd} \\[4pt]
\sfrac12 I_{(\ell-2)/2} & \textrm{for $\ell$ even}
\end{cases} \ .
\end{equation}

\end{appendix}

\bigskip

\small{

}

\end{document}